\documentclass[aps,pra,reprint,twocolumn,showpacs,superscriptaddress,groupedaddress]{revtex4-1}  
\usepackage{graphicx}  
\usepackage{dcolumn}   
\usepackage{bm}        
\usepackage{amssymb}   
\usepackage{amsmath}
\usepackage{xcolor}

\begin{document}



\title{Manipulation and exchange of Light with Orbital Angular Momentum in Quantum Dot Molecules}
\author{Mahboubeh Mahdavi}
\affiliation{Department of Physics, University of Zanjan, University Blvd., 45371-38791, Zanjan, Iran}
\author{Zahra Amini Sabegh}
\affiliation{Department of Physics, University of Zanjan, University Blvd., 45371-38791, Zanjan, Iran}
\author{Mohammad Mohammadi}
\affiliation{Department of Physics, University of Zanjan, University Blvd., 45371-38791, Zanjan, Iran}
\author{Hamid Reza Hamedi}
\email{hamid.hamedi@tfai.vu.lt}
\affiliation{Institute of Theoretical Physics and Astronomy, Vilnius University, Saul\.{e}tekio 3, Vilnius LT-10257, Lithuania}
\author{Mohammad Mahmoudi}
\email{mahmoudi@znu.ac.ir}
\affiliation{Department of Physics, University of Zanjan, University Blvd., 45371-38791, Zanjan, Iran\\}
\date{\today}

\begin{abstract}
We study the interaction of laser pulses carrying orbital angular momentum (OAM) with structural asymmetry quantum dot molecules characterized by four energy levels. We demonstrate how the inter-dot tunneling endows exchange of optical vortices between different frequencies. We consider a case where a weak probe beam has an optical vortex and thus has a zero intensity at the center. The presence of tunneling coupling generates an additional weak laser beam with the same vorticity as that of the incident vortex beam.  We analyze conditions for the vortex of the initial beam to be transferred efficiently to the generated beam. The propagation of Laguerre-Gaussian (LG) beams possessing OAM states characterized by both azimuthal and radial indices is then investigated for the case where the strong control beam is also an OAM mode. It is shown that the conservation of OAM states is always satisfied over the OAM exchange process. Yet, an abnormal case is observed in which the radial index induces some intensity patterns of the generated beam which differs from a pure LG beam of incident beams. Analytical solutions are provided to elucidate such effects induced by radial indices on propagation characteristics of OAM beams. When superimposing two initially present weak OAM modes, it is observed that the resulting optical vortices move about the beam axis as the light propagates, forming a sort of “constellation” around the center. The shift in axis of such a composite pulses is due to the effect of inter-dot tunneling which is controlled by an external electric voltage. The optical angular momenta may add a new degree of freedom in the study of solid systems suitable for quantum technologies.
\end{abstract}
\maketitle

\section{Introduction}
Quantum dots (QDs) are semiconductor nanoparticles possessing wide application in quantum optics and quantum information science because of their high nonlinear optical susceptibility, large electric-dipole moments of intersubband transitions, and great flexibility in designing devices \cite{Oost,Zrenner}. Their size is typically on the order of several nanometers in diameter \cite{Ashoori}. The electrons and holes in such a scale are confined in all three spatial dimensions making artificial atoms. An electron in such a molecule can pass via the inter-dot tunneling through potential barrier between quantum dots \cite{Oost}. The inter-dot tunneling induces the quantum coherence in quantum dot molecules (QDMs) which can be controlled by applying an external static gate voltage \cite{Villas,Muller}. It is well established that quantum interference can lead to various nonlinear optical phenomena in QDMs based on the inter-dot tunneling. Electromagnetically induced transparency based on the inter-dot tunneling has been introduced in $2006$ \cite{Zhu}. It has been demonstrated that the group velocity of the light pulse can be controlled by the electric voltage. The tunneling induced superluminal light propagation has been also investigated in QDMs \cite{Mahmoudi1}. Four-wave mixing optical response from a sample containing a non-homogenously broadened ensemble of vertically stacked pairs of quantum dots was theoretically studied \cite{Sitek}. It was demonstrated that the entanglement and quantum-information transfer between spatially separated QDMs can be controlled by the inter-dot tunneling \cite{Lu}. Several other works have also studied different optical phenomena in QDMs. Examples include optical bistability \cite{Mahmoudi2}, the transmission and reflection of pulse \cite{Mahmoudi3} and controlling the Goos-H\"{a}nchen shift \cite{Mahmoudi4}.

On the other hand, the orbital angular momentum (OAM) of light affixes a new degree of freedom to optical technologies enabling widespread applications in data transmission, optical communication \cite{Wang,Bozinovic}, optical tweezers \cite{Padgett}, and quantum information \cite{Molina}. The OAM light is a vortex beam which has a ring-shape intensity profile accompanied by the helical phase front \cite{Babiker}. Although having a history predating $1992$, Allen \textit{et al.} have been pioneers observing such a twisted light beams with helical wave fronts and a phase singularity that gives rise to a dark spot in the center with no intensity \cite{Allen}. Such lights with OAM can be created artificially through a variety of methods including cylindrical lens mode converters \cite{Beijersbergen}, spiral phase plates \cite{Beijersbergen2}, forked diffraction gratings \cite{Bazhenov}, computer-generated holograms \cite{Heckenberg}, and spatial light modulators (SLMs) \cite{Jesacher}. A number of interesting quantum optical effects appear when such a structured light interplays with the matter, such as the second harmonic generation (SHG) \cite{Dholakia,Courtial}, four-wave mixing (FWM) \cite{Barreiro,Walker,Persuy,Lanning}, sum-frequency generation \cite{Li}, and spatially structured electromagnetically induced transparency (EIT) \cite{Radwell,Hamedi1}. Recently and by using the Laguerre-Gaussian (LG) beams, the quantum entanglement between an ensemble of the three-level atomic systems and its spontaneous emissions has been investigated. It has been found that the atom-photon entanglement depends on the intensity profile as well as the OAM of the applied fields in the closed-loop atomic systems \cite{Amini}. The generation of an OAM-carrying ultra-violet (UV) light through SHG and OAM-entanglement frequency transducer has been experimentally examined by Zhou \textit{et al.} \cite{Zhou1,Zhou2}. Recent studies deal with the interaction of matter with twisted light and explore the possibility of exchange of optical vortices between different light frequencies \cite{Hamedi2,Hamedi3,Hamedi4,Hamedi5,Hong,Qiu,Arbiv,Ruseckas1,Ruseckas2,Amini2,Amini3}. Juzelin\={u}as team realized the exchange of OAM modes in four- and five-level quantum systems \cite{Ruseckas1,Ruseckas2}. Transfer of optical vortices has been shown to be possible in four-level EIT \cite{Hamedi2}, coherent population trapping (CPT) \cite{Hamedi3}, and phaseonium media \cite{Hamedi4}.

As we know, one can optically induce transitions between different electronic states in semiconductor QD nanostructures. Although plane waves (e.g., Gaussian laser beams) have mostly been employed to study the light-matter coupling, little work has been carried out to excite QDs with the OAM light. The transfer of energy from light to material is essential in quantum information processing \cite{Poynting}. The present work concentrates on the interaction of laser beams carrying OAM with QDMs. It is shown that due to the presence of the inter-dot tunneling, a single probe vortex beam initially acting on one transition of the four-level QDM generates an extra laser beam with the same vorticity as that of the incident vortex beam. Another favorable situation is then considered for the exchange of optical vortices in which the strong control beam represents also a LG beam. It is shown that the OAM number of the generated twisted beam stays conserved during the OAM transfer.  There exists an abnormal case, however, where the radial index develops some intensity-distribution patterns for the generated beam different from the initial beams. Analytical solutions are presented to explain this particular case.  If the two incident beams in LG modes are initially nonzero and are superimposed, they can generate a pattern of vortices with shifted axes once the beams are propagating inside the medium. Such a composite off-axis pattern of resulting beams is due to the effect of inter-dot tunneling which can be controlled by an external electric voltage.

The organization of the paper is as follows. In Section \ref{s2} we introduce the model and present the formulation of the basic set of equations by solving analytically the coupled Maxwell-Bloch equations. The results are presented in Section \ref{s3}, while Section \ref{s4} summarizes the main results.

\section{Theoretical framework}\label{s2}
\begin{figure}[b!]
\centering
\includegraphics[width=0.8\linewidth]{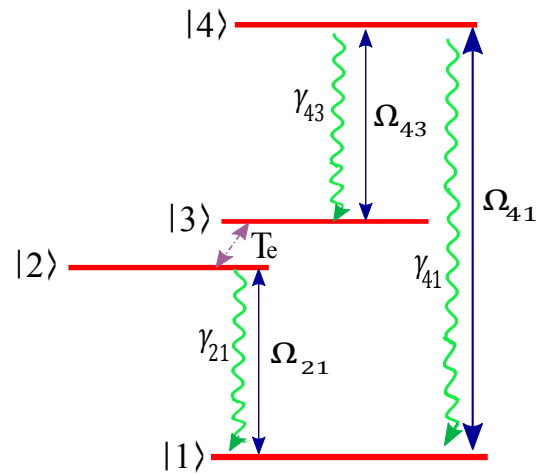}
\caption{ Schematic band structure and level configuration of the QDM system}
\label{f1}
\end{figure}

Let us consider a QDM consisting of two coupled QDs with different band structures coupled by tunneling, forming a four-energy-level double coupled QDM system (see Fig. \ref{f1}).
Such a QDM system can be fabricated using the self-assembled dot growth technology. As a realistic example, the asymmetric QDMs have been depicted in double-layer InAs/GaAs structure separated by a barrier \cite{Tarasov}. The transition $|i\rangle\leftrightarrow|j\rangle$ is excited by an external field with the frequency $\omega_{ij}$ and Rabi frequency $\Omega_{ij}=\vec{\mu}_{ij}\cdot\vec{E}_{ij}/\hbar (i,j\in1,..,4)$ where $\mu _{ij}$ and $E_{ij}$ are the induced dipole moment of the transition $|i\rangle\leftrightarrow|j\rangle$ and the amplitude of the applied field, respectively. A weak probe field is applied to the transition $|1\rangle\leftrightarrow|2\rangle$, while the transition $|3\rangle\leftrightarrow|4\rangle$ is excited by an strong control field. The states  $|2\rangle$ and $|3\rangle$ are coupled by the electron tunneling process. An extra weak field with a frequency $\omega_{41}=\omega_{21}+\omega_{32}+\omega_{43}$ is generated due to the three-wave mixing. Note that $\omega_{32}$ is considered to be zero owing to the negligible energy difference of the transition $|2\rangle\leftrightarrow|3\rangle$. Applying the dipole and rotating-wave approximations, the interaction Hamiltonian of the system can be written as
\begin{eqnarray}\label{e1}
&&H=\sum_{j} \varepsilon_{j}|j\rangle\langle j|+T_{e}|2\rangle\langle3|\nonumber\\ &&\qquad-\hslash[\Omega_{21}^{\ast}e^{i\Delta_{21}t}|1\rangle\langle2|+\Omega_{43}e^{-i\Delta_{43}t}|4\rangle\langle3|\nonumber\\ &&\qquad+\Omega_{4}^{\ast}e^{i\Delta_{41}t}|1\rangle\langle 4|+H.C.],
\end{eqnarray}
 where $ \varepsilon_{j}$ is the energy of the state $|j\rangle$ and $\Delta_{ij}=\omega_{ij}-\overline{\omega}_{ij}$ describes the frequency detuning between the applied laser field and resonant frequency,  associating with the corresponding transitions, $|i\rangle\leftrightarrow|j\rangle$. The parameter $T_{e}$ represents the strength of the inter-dot tunneling between the states $|2\rangle$ and $|3\rangle$ created by a static electric field. The density-matrix equations for the matter fields are
 \begin{eqnarray}\label{e2}
  \dot{\rho}_{11}&=&2\Gamma_{21}\rho_{22}+2\Gamma_{41}\rho_{44}+i\Omega^{\ast}_{21}\rho_{21}-i\Omega_{21}\rho_{12}+i\Omega^{\ast}_{41}\rho_{41}\nonumber\\
  &-&i\Omega_{41}\rho_{14},\nonumber\\
  \dot{\rho}_{22}&=&-2\Gamma_{21}\rho_{22}+i\Omega_{21}\rho_{12}-i\Omega^{\ast}_{21}\rho_{21}
  +iTe(\rho_{32}-\rho_{23}),\nonumber\\
  \dot{\rho}_{33}&=&2\Gamma_{43}\rho_{44}+i\Omega^{\ast}_{43}\rho_{43}-i\Omega_{43}\rho_{34}
  +iTe(\rho_{23}-\rho_{32}),\nonumber\\
  \dot{\rho}_{12}&=&-(i\Delta_{21}+\Gamma_{21})\rho_{12}+i\Omega_{21}(\rho_{22}-\rho_{11})-iTe\rho_{13}\nonumber\\
  &+&i\Omega^{\ast}_{41}\rho_{42},\nonumber\\
  \dot{\rho}_{13}&=&-i[\Delta_{41}-\Delta_{43}]\rho_{13}-i\Omega_{43}\rho_{14}+i\Omega^{\ast}_{21}\rho_{23}+i\Omega^{\ast}_{41}\rho_{43}\nonumber\\
  &-&iTe\rho_{12},\nonumber\\
  \dot{\rho}_{14}&=&-[i\Delta_{41}+\Gamma_{41}]\rho_{14}-i\Omega^{\ast}_{43}\rho_{13}+i\Omega_{41}(\rho_{44}-\rho_{11})\nonumber\\
  &+&i\Omega^{\ast}_{21}\rho_{24},\nonumber\\
  \dot{\rho}_{23}&=&-\Gamma_{21}\rho_{23}-i[\Delta_{41}-\Delta_{21}-\Delta_{43}]\rho_{23}+i\Omega_{21}\rho_{13}\nonumber\\
  &-&i\Omega_{43}\rho_{24}+iTe(\rho_{33}-\rho_{22}),\nonumber\\
   \dot{\rho}_{24}&=&-[(i(\Delta_{41}-\Delta_{21})+\Gamma_{21}+\Gamma_{43}+\Gamma_{41})]\rho_{24}+iTe\rho_{34}\nonumber\\
  &+&i\Omega_{21}\rho_{14}-i\Omega^{\ast}_{43}\rho_{23}-i\Omega^{\ast}_{41}\rho_{21},\nonumber\\
  \dot{\rho}_{34}&=&-[i\Delta_{43}+(\Gamma_{41}+\Gamma_{43})]\rho_{34}+iTe\rho_{24}-i\Omega^{\ast}_{41}\rho_{31}\nonumber\\
  &+&i\Omega^{\ast}_{43}(\rho_{44}-\rho_{33}),\nonumber\\
  \dot{\rho}_{44}&=&-(\dot{\rho}_{11}+\dot{\rho}_{22}+\dot{\rho}_{33}).
  \end{eqnarray}
The above density matrix equations represent the evolution of the system affected by the laser fields and tunneling coupling. They follow from the general quantum Liouville equation for the density matrix operator
\begin{equation}\label{e3}
\frac{\partial\rho}{\partial t}=\frac{-i}{\hbar}[H,\rho]+L(\rho),
\end{equation}
where the damping operator $L(\rho)$ describes the decoherence processes.
The steady state analytical expressions for the coherence terms $\rho_{21}$ and $\rho_{41}$  can be obtained, by solving Eq. (\ref{e2}) for $\Gamma_{21}=\Gamma_{41}=\Gamma_{43}=\gamma$ under multi-photon resonance condition, $\Delta_{43}=0$, $\Delta_{21}=\Delta_{41}=\Delta$, giving
\begin{flalign}\label{e4}
\rho_{21}&=\frac{(T_{e}\Omega_{41}-\Omega_{43}\Omega_{21})\Omega_{43}^{\ast}+i(\gamma\Delta-\Delta^{2})\Omega_{21}}{(\gamma-i\Delta)(iT_{e}^{2}+\gamma\Delta-i\Delta^{2}+i|\Omega_{43}|^{2})},\nonumber\\
\rho_{41}&=-\frac{T_{e}(T_{e}\Omega_{41}-\Omega_{43}\Omega_{21})-i\gamma\Delta\Omega_{41}-\Delta^{2}\Omega_{41}}{(\gamma-i\Delta)(iT_{e}^{2}+\gamma\Delta-i\Delta^{2}+i|\Omega_{43}|^{2})}.
\end{flalign}
 The Maxwell wave equations in the slowly varying envelope approximation read
\begin{eqnarray}\label{e5}
\frac{\partial\Omega_{21}(z)}{\partial z}&=&i\frac{\alpha\gamma}{2L}\rho_{21},\nonumber\\
\frac{\partial\Omega_{41}(z)}{\partial z}&=&i\frac{\alpha\gamma}{2L}\rho_{41},
\end{eqnarray}
where $L$ and $\alpha$ are the length of the QDM ensemble and the optical depth for both fields \cite{Hamedi2}, respectively. Substituting Eq. (\ref{e4}) into Eq. (\ref{e5}) and assuming $\Delta=0$, $\Omega_{41}(z=0)=0$ and $\Omega_{21}(z=0)=\Omega_{21}$, one arrives at the following equations
\begin{flalign}\label{e6}
\Omega_{21}(r,\varphi,z)=\frac{\Omega_{21}(r,\varphi)(T_{e}^{2}+\exp(-\frac{z\alpha}{2L})|\Omega_{43}|^{2})}{T_{e}^{2}+|\Omega_{43}|^{2}},
\end{flalign}
and
\begin{flalign}\label{e7}
\Omega_{41}(r,\varphi,z)=-\frac{(-1+\exp(-\frac{z\alpha}{2L})) T_{e}\Omega_{21}(r,\varphi)\Omega_{43}}{T_{e}^{2}+|\Omega_{43}|^{2}},
\end{flalign}
which describe the propagation of fields inside the medium.
\section{Results and discussions}\label{s3}
The complex form describing the distribution of the field amplitude of a LG beam can be expressed cylindrically as

\begin{eqnarray}\label{e8}
\Omega(r,\varphi)&=&\Omega_{0}\frac{1}{\sqrt{|l|!}}(\frac{\sqrt{2}r}{w_{LG}})^{|l|}\nonumber\\
&&\times L_{p}^{|l|}(2r^{2}/w_{LG}^{2})~e^{-r^{2}/w_{LG}^{2}}~e^{il\varphi},
\end{eqnarray}
where $\Omega_{0}$, $w_{LG}$, $l$, and $p$ show the constant Rabi frequency, beam waist radius, azimuthal (OAM) and radial indices of the LG modes, respectively. Here, the associated Laguerre polynomial, $L_{p}^{|l|}$ has the form
\begin{equation}\label{e9}
    L_{p}^{|l|}(x)=\frac{e^{x}x^{-|l|}}{p!} \frac{d^{p}}{dx^{p}}[x^{|l|+p}e^{-x}],
\end{equation}
with $x=2r^{2}/w_{LG}^{2}$ determining the radial dependence of the LG beams for different radial mode numbers.  When  $l$  is not zero, the LG light beams possess OAM along the optical axis.

\subsection{Exchange of vortices}

Let us now consider the spatial profile of the laser fields described by Eq. (\ref{e8}) . As Eq. (\ref{e7}) shows, the Rabi frequency of the generated third field corresponds to the inter-dot tunneling as well as the Rabi frequency of the probe and the strong control fields. Thus, the generated laser field $\Omega_{41}$ is a vortex if any of the fields $\Omega_{21}$, $\Omega_{43}$  or both of them are initially vortices.
Such a transfer of optical vortices is because of the presence of the inter-dot tunneling which can be controlled by applying an external electric voltage.

\begin{figure}[htbp]
\centering
  \includegraphics[width=1.0\linewidth]{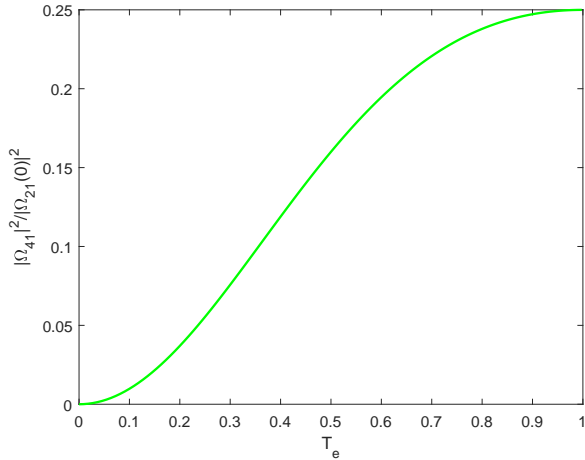}
  \caption{\small  The dimensionless intensity of the generated third field $|\Omega_{41}(z)|^{2}/|\Omega_{21}(0)|^{2}$ versus the tunneling strength for $z=L$, $\Omega_{43}=\gamma$, and $\alpha=20$.}\label{f2}
\end{figure}

Let us first assume that  only the probe field $\Omega_{21}$ is vortex. The effect of the inter-dot tunneling strength, $T_{e}$, on the dimensionless intensity of the generated  field $|\Omega_{41}(z)|^{2}/|\Omega_{21}(0)|^{2}$ has been shown by means of Eq. (\ref{e7}). The dimensionless plot for the intensity of the generated OAM field as a function of the tunneling strength is shown at $z=L$ in Fig. \ref{f2} for $w_{LG}=0.5 mm$, $\Omega_{43}=\gamma$, and $\alpha=20$. As expected, the intensity of the generated third field is zero when the inter-dot tunneling strength is zero $T_{e}=0$. The generated third field grabs its maximal value for $T_{e}=1$. This is an optimal value of tunneling coupling for which the intensity of the generated OAM mode is the largest.
\begin{figure}[htbp]
\centering
  \includegraphics[width=1.0\linewidth]{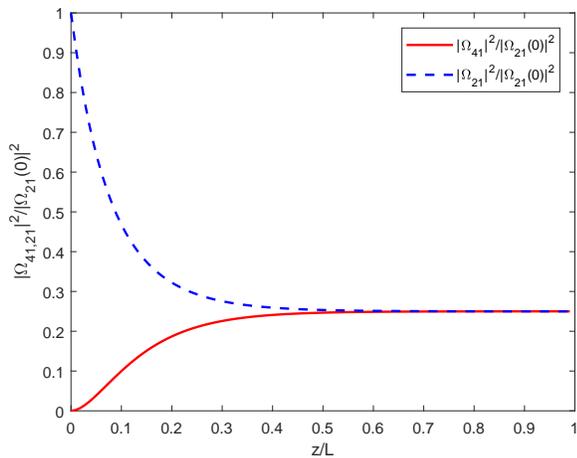}
  \caption{\small The  dimensionless intensity of fields $|\Omega_{21}(z)|^{2}/|\Omega_{21}(0)|^{2}$ and $|\Omega_{41}(z)|^{2}/|\Omega_{21}(0)|^{2}$ versus the dimensionless distance $z/L$ for $T_{e}=1$. Other parameters are the same as in Fig. \ref{f2}.}\label{f3}
\end{figure}

Next we consider the optimal inter-dot tunneling effect and investigate the intensity of the probe and generated third beams inside QDM medium, using Eqs. (\ref{e6}) and (\ref{e7}). Figure  \ref{f3} shows the dimensionless intensities $|\Omega_{21}(z)|^{2}/|\Omega_{21}(0)|^{2}$ and  $|\Omega_{41}(z)|^{2}/|\Omega_{21}(0)|^{2}$  against the dimensionless distance $z/L$ for $T_{e}=1$. The other parameters are the same as in Fig. \ref{f2}. One can see that the laser field $\Omega_{41}$ has not yet been created at the beginning of the ensemble where the weak probe beam has just entered. Propagating inside the QD ensemble, the beam $\Omega_{41}$  is generated. Equations (\ref{e6}) and (\ref{e7}) and Fig. \ref{f3} indicate that both OAM beams experience energy losses mostly at the beginning of the ensemble, going deeper into the ensemble losses disappear where the system is transferred to some transparency state (see Fig. \ref{f3}).
\begin{figure*}[htbp]
\centering
  \includegraphics[width=1.0\linewidth]{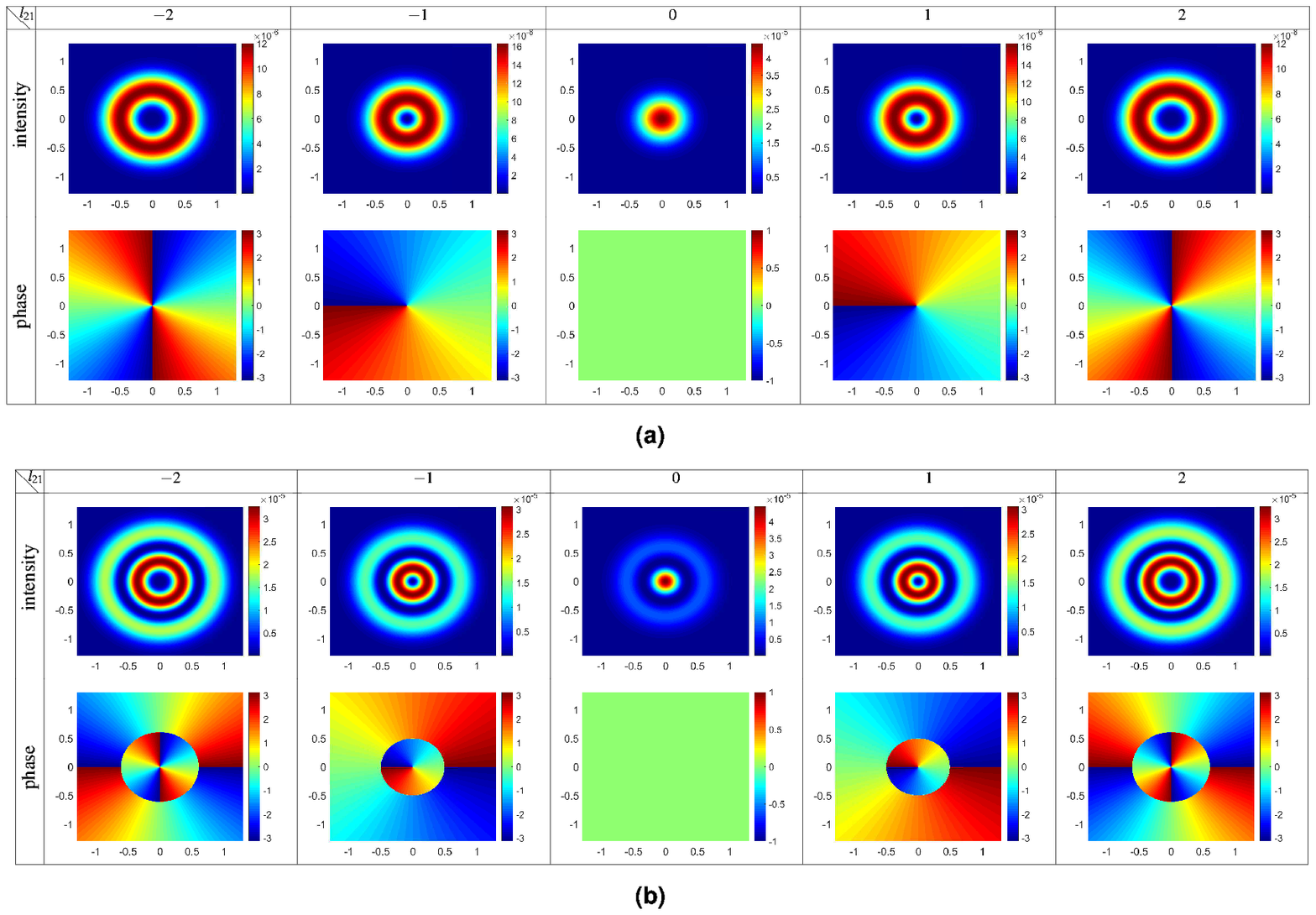}
  \caption{\small The intensity and phase profiles of the generated third field as a function of $x$ and $y$ for the different modes of the probe LG field, $l_{21}=-2,-1,..,2$, with $p_{21}=0$ (a) and $p_{21}=1$ (b). The inter-dot tunneling parameter, beam waist and constant Rabi frequency of the probe LG field are chosen as $T_{e}=1$, $w_{LG}=0.5 mm$, and $\Omega_{21_{0}}=0.01\gamma$, respectively. The value of the inter-dot tunneling parameter is, $T_{e}=1$. Other parameters are the same as in Fig. \ref{f2}.}\label{f4}
\end{figure*}

In what follows, we proceed with the numerical simulations to describe such a swapping in OAM modes based on the inter-dot tunneling effect.

\subsubsection{First case: Only $\Omega_{21}$ is a vortex}

 In Fig. \ref{f4}(a), we show the intensity and phase profiles of the generated OAM mode as a function of $x$ and $y$ for different winding numbers $l_{21}=-2,-1,..,2$ but with zero radial index $p_{21}=0$. The horizontal and vertical axes $x$ and $y$ are scaled in $mm$. We take $T_{e}=1$, $w_{LG}=0.5 mm$, and $\Omega_{21_{0}}=0.01\gamma$, and the other parameters the same as in Fig. \ref{f2}. It is observed that intensity profile of the third generated field has a Gaussian profile when $l_{21}=0$. Yet, the doughnut intensity profiles appear with a dark (blue) hollow center for nonzero $l_{21}$, indicating a conserved transfer of optical vortex of the probe beam to the generated third beam. The diameter of the doughnuts increases for the larger topological charges $l_{21}$. The helical phase patterns help to realize the nature of the singularity at the core of the generated third OAM beam. No singularity takes place at phase patterns when $l_{21}=0$ confirming a Gaussian-shaped wavefront of the laser field with a normal phase. The phase patterns start twisting for nonzero vorticities.

Note that all profiles shown in Fig. \ref{f4}(a) exhibit a single ring in their intensity patterns, indicating their radial indices are all at zero. Let us study on Fig. \ref{f4}(b), the effect of the nonzero radial index $p$ ($p_{21}=1$) on the intensity and phase profiles of the generated OAM field for the different azimuthal indices $l_{21}=-2,-1,..,2$.
While the central dark holes always exist for non-zero vorticities, the radial index of the probe LG beam develops some remarkable changes in the intensity and phase profiles of the generated third field. In particular, there exists now a dark ring between two bright rings in each diagram for the intensity profile. However, measuring an intensity profile is not always a very accurate way to determine different mode indices for a particular vortex beam, as sometimes the bright rings are too dim to be distinguished by the naked eye (e.g., see intensity diagrams shown later in Fig. \ref{f5}). Helical phase profiles, instead, provide an accurate and convenient way to check for different azimuthal and radial indices. As an example, let us read the case with $l_{21}=2$ and $p_{21}=1$, illustrated in the last diagram in the second row of Fig. \ref{f4}(b). The phase jumps from 0 to $2\times 2\pi=4\pi$ at the beam center, indicating an azimuthal index $l_{21}$ equal to 2.  Two zones appear from the core to the border of the phase profile, at their boundary the phase diagram experiences a $\pi$-shift, where the corresponding light goes to zero in intensity. The radial index for the generated beam is read to be 1, which is due to the existence of two zones in the profile separated by one $\pi$-shift boundary circle. To read and identify $l$ and $p$ numbers for any unknown vortex beam, one can develop a general manner; the azimuthal index is read with $n l$ if the phase jumps from 0 to 2$n\pi$ around the singularity point, while the radial index is read with $m p$ if the phase diagram demonstrates a number of $m\pi$-shift boundary circles from the radial direction. One can use the above manner to distinguish different azimuthal and radial indices for more complex vortex beams such as those described in the next sections.

\subsubsection{Second case: Both $\Omega_{21}$ and $\Omega_{43}$ are vortices}
\begin{figure}[htbp]
\centering
  \includegraphics[width=1.0\linewidth]{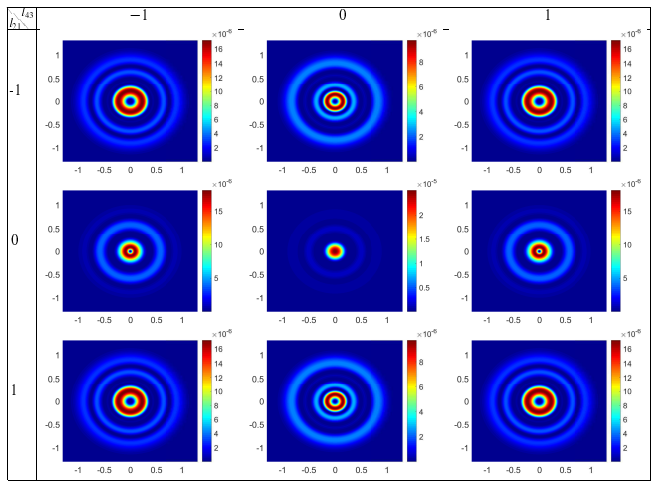}
  \caption{\small The intensity profiles of the generated field versus $x$ and $y$ for different modes of the weak probe and strong control LG fields with $p_{21}=1$, $p_{43}=2$, $l_{21},l_{43}\in-1,0,1$, $\Omega_{21_{0}}=0.01\gamma$, $\Omega_{43_{0}}=\gamma$. Other parameters are the same as in Fig. \ref{f4} }\label{f5}
\end{figure}

\begin{figure}[htbp]
\centering
  \includegraphics[width=1.0\linewidth]{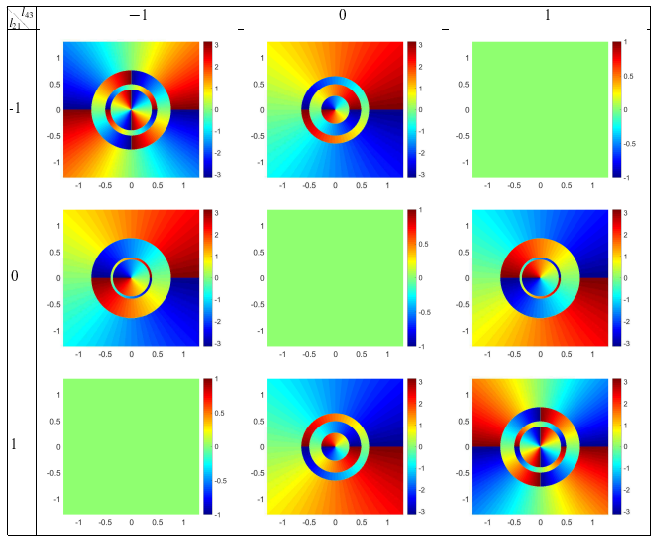}
  \caption{\small The phase profiles of the generated third field versus $x$ and $y$ for the different modes of the applied LG fields. The parameters used here are the same as in Fig. \ref{f5}.}\label{f6}
\end{figure}

Next, we study a situation where both $\Omega_{21}$ and $\Omega_{43}$  describe optical vortices. In Fig. \ref{f5}, we depict the intensity of the generated third field as a function of $x$ and $y$ for different modes of the probe and strong control LG beams with $p_{21}=1$, $p_{43}=2$, $l_{21},l_{43} \in-1,0,1$, $\Omega_{21_{0}}=0.01\gamma$ and $\Omega_{43_{0}}=\gamma$, while the other parameters are the same as in Fig. \ref{f4}. Figure \ref{f6} shows the corresponding phase patterns.
It is seen that the generated third beam develops a new OAM state with the total azimuthal index of $l_{21}+l_{43}$, meaning that the total azimuthal index of applied beams keeps still constant over the OAM exchange process. This is quite understandable from  Eq. (\ref{e7}) as it is as a result of the OAM conservation. In a particular case when $l_{21}=-l_{43}\neq0$, the generated third beam is no longer a vortex as the total topological charge associated with the sum of incident beams becomes zero. As an example, let us consider the last diagrams in the first row in Figs. \ref{f5} and \ref{f6}. As expected, the phase diagram of the generated beam shows no singularity at the core. However, the intensity profile demonstrates a multi-ring pattern. Note that such a pattern is not called a vortex, although it has a zero intensity at the beam center. This abnormal intensity-distribution is due to the non-zero radial index for the incident beams  $\Omega_{21}$ and $\Omega_{43}$.
According to Eqs. (\ref{e8}) and (\ref{e9}), the radial coordinate dependence of the LG beam is modified when  $p>0$, resulting in $p+1$ concentric rings in the intensity profiles with zero-intensity center.

\begin{figure*}[t!]
\centering
  \includegraphics[width=1.0\linewidth]{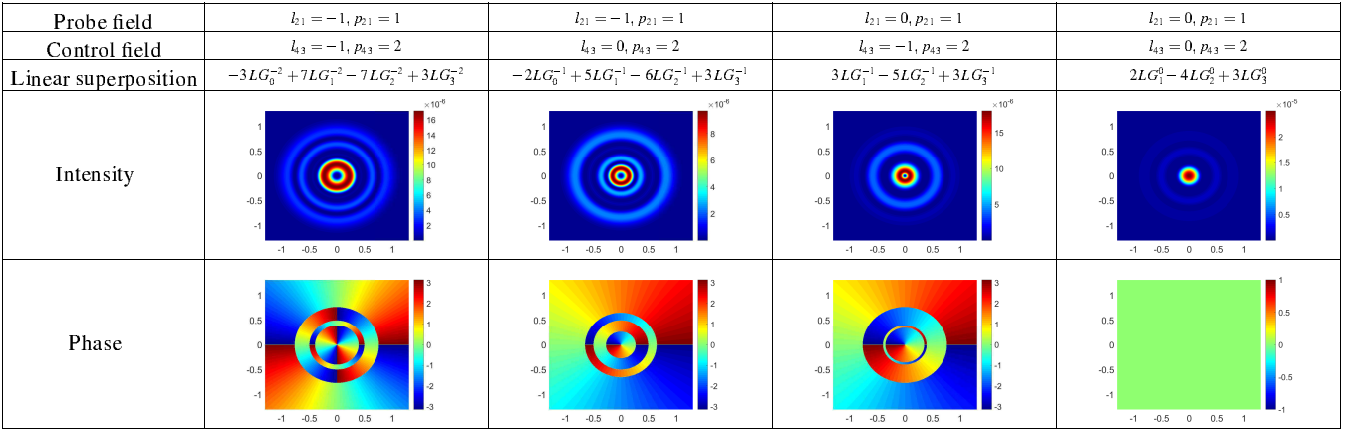}
  \caption{\small The intensity and phase profiles of the corresponding superposition states by considering the effect of the QDM medium.}\label{f7}
\end{figure*}

 According to Eq. (\ref{e7}), the generated third field is proportional to the product of the weak probe $\Omega_{21}$ and strong control $\Omega_{43}$ fields which are considered as the LG modes. When $p=0$, the product of two Rabi frequency of the LG modes makes a single mode with $l=l_{21}+l_{41}$, but the situation is completely different for $p>0$. The product of two LG modes can be expanded as a linear superposition of different LG modes. The appropriate superposition of the contributing LG modes of the generated third field has been analytically obtained for several different LG modes. The detail of calculation is given in the Appendix A when $l_{21}=-1,p_{21}=1$ and $l_{41}=-1,p_{41}=2$. The coefficients appeared in the superposition mode determine the contribution of each LG mode to generate the third field. The intensity and phase profiles of the corresponding linear superpositions are shown in Fig. \ref{f7} for several different LG modes. The first and second rows show the characteristics of the weak probe and strong control fields, respectively, while the third row illustrates the linear superposition for different LG modes forming the generated third field. Note that such a linear superposition state of different LG modes takes place only under the effect of the QDM medium. Therefore,  the transfer of OAMs is accompanied by the exchange of different radial modes of the applied LG fields.
\subsection{Composite vortices}
\begin{figure}
\centering
  \includegraphics[width=1.0\linewidth]{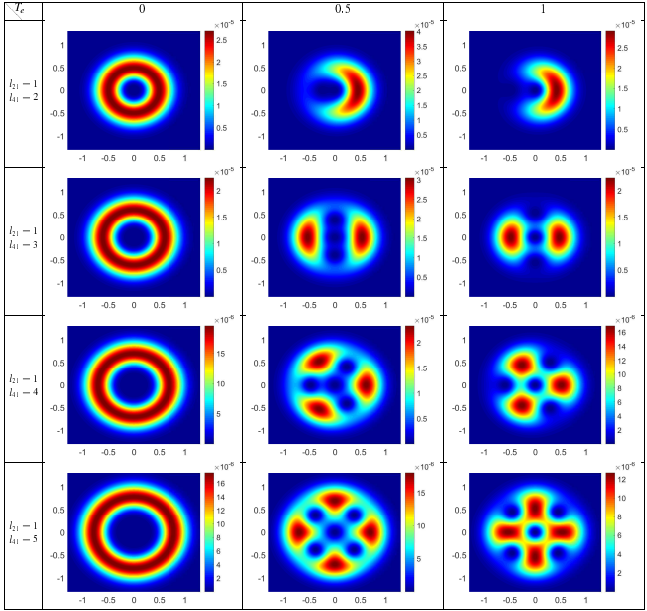}
  \caption{\small The intensity profiles of the third field versus $x$ and $y$ for the different modes of the probe LG fields, i.e., ($l _{21}=1, l _{41}=2$), ($l _{21}=1, l _{41}=3$), ($l _{21}=1, l _{41}=4$) and ($l _{21}=1, l _{41}=5$), with  $p _{21}= p _{41}=0$ for different inter-dot tunneling parameters, $T _{e}=0, 0.5, $ and $1$ .The constant Rabi frequency of the third field is chosen as $\Omega_{41_{0}}=0.01\gamma$ and the other parameters are the same as Fig. \ref{f2}.}\label{f8}
\end{figure}
\begin{figure}[htbp]
\centering
  \includegraphics[width=1.0\linewidth]{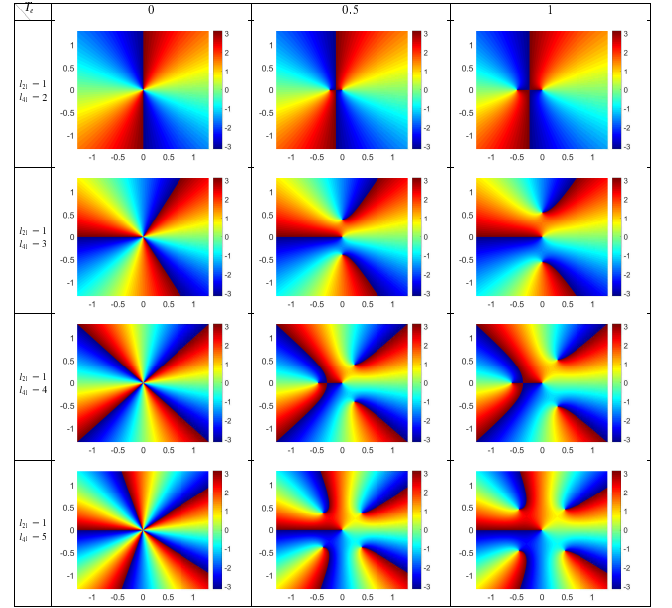}
  \caption{\small The phase profiles of the third field versus $x$ and $y$. The parameters used here are the same as Fig. \ref{f8}.}\label{f9}
\end{figure}

Let us next consider a case where both vortex beams $\Omega_{41}(z=0)=\Omega_{41}(r,\varphi)$ and $\Omega_{21}(z=0)=\Omega_{21}(r,\varphi)$  are incident on the medium. The analytical expression describing the propagation of $\Omega_{41}$ then takes the form
\begin{flalign}\label{e10}
\Omega_{41}(r,\varphi,z)&=\frac{(\exp(-\frac{z\alpha}{2L})T_{e}^{2}+|\Omega_{43}|^{2})\Omega_{41}(r,\varphi)}{T_{e}^{2}+|\Omega_{43}|^{2}}\nonumber\\ &+\frac{(1-\exp(-\frac{z\alpha}{2L}))T_{e}\Omega_{43}\Omega_{21}(r,\varphi)}{T_{e}^{2}+|\Omega_{43}|^{2}}.
\end{flalign}
The intensity and helical phase profiles of the generated field $\Omega_{41}$ are displayed in Figs. \ref{f8} and \ref{f9} for different inter-dot tunneling parameters, $T _{e}=0, 0.5, $ and $1$. Different modes of the probe LG fields are considered, i.e., ($l _{21}=1, l _{41}=2$), ($l _{21}=1, l _{41}=3$), ($l _{21}=1, l _{41}=4$) and ($l _{21}=1, l _{41}=5$), while  $p _{21}= p _{41}=0$. We take $\Omega_{41_{0}}=0.01\gamma$ and the other parameters are the same as Fig. \ref{f2}. The left column of Fig. \ref{f8} shows that for $T_{e}$=0, the two incident LG fields, $\Omega_{41}(0)$ and $\Omega_{21}(0)$, do not interact with each other; hence, the output field featured by Eq. (\ref{e10}) contains the same vorticity as the input field $\Omega_{41}(0)$. The incident applied LG fields start interacting in the presence of the inter-dot tunneling effect, forming the composite vortices. The tunneling coupling grows more singularities at the transverse plane associated with zero-intensity regions. For instance, when $T_{e}=0.5$ (as indicated in the middle columns), the resulting composite beam $\Omega_{41}$ exhibits a singularity at the core surrounded by some peripheral vortices (constellation patterns). Increasing the inter-dot tunneling parameter to $T_{e}=1$, moves the position of peripheral vortices far away from the central vortex. Generally speaking, if $|l _{21}|<|l _{41}|$, the resulting composite beam acquires a vortex of vorticity $|l _{21}|$ located at the beam core which is surrounded by $|l _{21}-l _{41}|$ peripheral vortices.
\section{Concluding Remarks}\label{s4}
To conclude, we have investigated the interplay of light beams with OAM with structural asymmetry quantum dot molecules with four energy levels. It has been shown that the inter-dot tunneling effect can induce the OAM transfer between different frequencies. We have considered a particular situation where a weak probe beam is initially a vortex beam. Due to the effect of tunneling coupling, an extra laser is generated with the same winding number as that of the incident probe field. An efficient condition is considered for such exchange of optical vortices. We have also studied the propagation of LG beams with azimuthal and radial indices when the strong control beam is also an OAM mode. An abnormal case is observed in which the radial index induces some intensity patterns for the generated beam which is quite different from the incident pure LG beams. An analytical model is presented to understand such an effect of radial index. We have also shown that when the two vortex beams are present at the beginning of the medium and as a result of tunneling coupling, composite vortices can take place with shifted axes.
\begin{acknowledgments}
M. Mahmoudi would like to express the deepest appreciation to Prof. G. Juzeli\={u}nas for useful discussion.  The research was supported in part by Iran's National Elites Foundation (INEF) (Shahid Chamran's Scientific Prize, Grant No. 15/10597).
\end{acknowledgments}

\appendix

\section{}

In this appendix, we evaluate the pure LG modes that contributed to the generation of the third field. Considering two applied fields as
\begin{eqnarray}\label{A1}
 \Omega_{21}(r,\varphi)&=&\Omega_{21_{0}}\frac{1}{\sqrt{|l_{21}|!}}(\frac{\sqrt{2}r}{w_{LG}})^{|l_{21}|}e^{-il_{21}\varphi}\nonumber\\
 &&\times L_{1}^{|l_{21}|}(2r^{2}/w_{LG}^{2})~e^{-r^{2}/w_{LG}^{2}},\nonumber\\
 \Omega_{43}(r,\varphi)&=&\Omega_{43_{0}}\frac{1}{\sqrt{|l_{43}|!}}(\frac{\sqrt{2}r}{w_{LG}})^{|l_{43}|}e^{-il_{43}\varphi}\nonumber\\
 &&\times L_{2}^{|l_{43}|}(2r^{2}/w_{LG}^{2})e^{-r^{2}/w_{LG}^{2}},
\end{eqnarray}
with $l_{21}=l_{43}=-1$, the generated third field has the form
\begin{eqnarray}\label{A2}
 &&\Omega_{41}(r,\varphi,z)=-\frac{(-1+\exp(-\frac{z\alpha}{2L})) T_{e}\Omega_{21}\Omega_{43}}{T_{e}^{2}+|\Omega_{43}|^{2}}\nonumber\\
 &&\qquad\qquad\quad=-\frac{(-1+\exp(-\frac{z\alpha}{2L})) T_{e}}
 {T_{e}^{2}+|\Omega_{43}|^{2}}\Omega_{21_{0}}\Omega_{43_{0}}\nonumber\\
 &&\qquad\qquad\qquad \times(\frac{1}{\sqrt{|-1|!}})^{2}((\frac{\sqrt{2}r}
 {w_{LG}})^{|-1|})^{2}\nonumber\\
 &&\qquad\qquad\qquad \times e^{-2r^{2}/w_{LG}^{2}}e^{-2i\varphi}L_{1}^{|-1|}(2r^{2}/w_{LG}^{2})\nonumber\\
 &&\qquad\qquad\qquad \times L_{2}^{|-1|}(2r^{2}/w_{LG}^{2}).
\end{eqnarray}
Some of the associated Laguerre polynomials can be used for obtaining the superposition of the pure LG modes forming the generated third field. The needed associated Laguerre polynomials are
\begin{eqnarray}\label{A3}
 &&L_{1}^{|-1|}(x)=(2-x),\nonumber\\
 &&L_{2}^{|-1|}(x)=\frac{1}{2}(6-6x+x^{2}),\nonumber\\
 &&L_{0}^{|-2|}(x)=1,\nonumber\\
 &&L_{1}^{|-2|}(x)=3-x,\nonumber\\
 &&L_{2}^{|-2|}(x)=\frac{1}{2}(12-8x+x^{2}),\nonumber\\
 &&L_{3}^{|-2|}(x)=\frac{1}{6}(60-60x+15x^{2}-x^{3}).
\end{eqnarray}
Substituting Eq. (\ref{A3}) in Eq. (\ref{A2}), the following superposition of pure LG modes is obtained for the generated third field
\begin{widetext}
\begin{eqnarray}\label{A4}
 &&\Omega_{41}(r,\varphi,z)=-\frac{(-1+\exp(-\frac{z\alpha}{2L})) T_{e}}
 {T_{e}^{2}+|\Omega_{43}|^{2}}\Omega_{21_{0}}\Omega_{43_{0}}(\frac{1}{\sqrt{|-1|!}})^{2}((\frac{\sqrt{2}r}
 {w_{LG}})^{|-1|})^{2} e^{-2r^{2}/w_{LG}^{2}}e^{-2i\varphi}
 (6-9x+4x^{2}-\frac{1}{2}x^{3}),\nonumber\\
&&\qquad\qquad\quad=-\frac{(-1+\exp(-\frac{z\alpha}{2L})) T_{e}}
 {T_{e}^{2}+|\Omega_{43}|^{2}}\Omega_{21_{0}}\Omega_{43_{0}}e^{-r^{2}/w_{LG}^{2}}
(-3LG_{0}^{-2}+7LG_{1}^{-2}-7LG_{2}^{-2}+3LG_{3}^{-2}),
\end{eqnarray}
\end{widetext}
in which $LG_p^l$ indicates a pure LG mode with the radial index $p$ and azimuthal index $l$.

\bibliographystyle{}

\end{document}